\input{aipcheck}

\documentclass[
    ,final            
  ]
  {aipproc}

\layoutstyle{6x9}

\begin{document}

\title{Preliminary Results from the Caltech Core-Collapse Project (CCCP)}


\author{Avishay Gal-Yam}{
  address={Division of Physics, Mathematics and Astronomy, California Institute of Technology}
  ,altaddress={Hubble Fellow} 
}

\author{S. B. Cenko}{
  address={Division of Physics, Mathematics and Astronomy, California Institute of Technology}
}

\author{D. B. Fox}{
  address={Department of Astronomy and Astrophysics, Pennsylvania State University}
}

\author{D. C. Leonard}{
  address={Department of Astronomy, San Diego State University}
}

\author{D.-S. Moon}{
  address={Division of Physics, Mathematics and Astronomy, California Institute of Technology}
}

\author{D. J. Sand}{
  address={Steward Observatory, University of Arizona}
}

\author{A. M. Soderberg}{
  address={Division of Physics, Mathematics and Astronomy, California Institute of Technology}
}

\begin{abstract}

We present preliminary results from the Caltech Core-Collapse
Project (CCCP), a large observational program 
focused on the study of core-collapse SNe. Uniform, high-quality NIR and optical photometry
and multi-epoch optical spectroscopy have been obtained using the 200'' Hale and robotic 60''
telescopes at Palomar, for a sample of 50 nearby core-collapse SNe. The combination of both 
well-sampled optical light curves and multi-epoch spectroscopy will enable spectroscopically 
and photometrically based subtype definitions to be disentangled from each other. Multi-epoch spectroscopy
is crucial to identify transition events that evolve among subtypes with time. The CCCP SN sample
includes every core-collapse SN discovered between July 2004 and September 2005 that was visible from
Palomar, found shortly ($< 30$ days) after explosion (based on available pre-explosion
photometry), and closer than $\sim120$ Mpc. This complete sample allows, for the first time, a
study of core-collapse SNe as a population, rather than as individual events. Here, we present
the full CCCP SN sample and show exemplary data collected. We analyze available data for the first 
$\sim 1/3$ of the sample and determine the subtypes of 13 SNe II based on both light curve shapes
and spectroscopy. We discuss the relative SN II subtype fractions in the context of associating 
SN subtypes with specific progenitor stars.

\end{abstract}

\maketitle


\section{Introduction}

Supernovae (SNe) play an important role in almost all areas of astrophysical research.
Obviously central to stellar evolution, these explosions also trigger (and inhibit) star formation,
produce heavy elements, dust and cosmic rays, and their energy input to the ISM is
a crucial ingredient in galaxy formation. Supernovae are probably involved in the formation
of neutron stars, black holes, and gamma-ray bursts (GRBs), and, as precision distance
estimators, were instrumental in establishing the present cosmological standard model.

We recognize two physically defined classes: 
thermonuclear SNe (type Ia), occurring when a white dwarf (WD) star is pushed over
the Chandrasekhar limit due to accretion from, or a merger with, a binary companion; and
all other types of SNe (II, Ib, Ic) resulting from gravitational core-collapse of massive
stars. The basic physical process of core-collapse SNe is known: a massive star (directly observed
in a few nearby cases; Fig. 1) undergoes gravitational core-collapse forming
a compact remnant. A mechanical shock, possibly assisted by neutrino energy deposition
(e.g., the neutrino burst detected from SN 1987A), is launched, and ejects the envelope of the
star.

However, despite this basic understanding, many fundamental questions remain. The
continued failure of the spherical neutrino-driven shock mechanism to actually explode massive
stars in 2- and 3-d numerical simulations, along with the advent of models invoking
bipolar or jet-driven core-collapse induced explosions (Khokhlov et al. 1999; Burrows et
al. 2006) has reopened the explosion mechanism question for core-collapse SNe. Different
mechanisms predict unique mappings between progenitors and the resulting SNe (e.g., Heger
et al. 2003), but we currently lack the corresponding observational basis to relate measured
SN properties (brightness, light curve shape, spectroscopic abundances of H, He and other
elements) to those of the putative progenitor (age, initial mass, mass loss history, binarity),
and thus constrain explosion models. The sum of our current knowledge is presented
in Fig. 1. As can be seen there, a robust link between progenitor classes and SN types
requires additional work. If the impact of the first progenitor-SN link (the blue supergiant
progenitor of SN 1987A) is any indication, as the progenitor-SN map becomes more robust, it
should provide a key input to SN models, and drive toward better understanding of SN physics.
An additional benefit is that linking specific
massive progenitor stars with different SN subtypes, in tandem with measurement of the
rate and relative frequency and properties of subtypes (II-P, IIn, Ib, Ic, etc.)
would ultimately allow to directly measure the stellar IMF out to high redshift (via
detection of SNe, much brighter than any single star), and to calculate
the role of core-collapse SNe in metal enrichment and energetics of the ISM, 
a prime topic of study with JWST, LSST, and the next generation of large telescopes.

With the goal of producing a robust progenitor-SN map (Fig. 1) to be used as a key
to better understand core-collapse SN physics and the IMF, we are pursuing two parallel
efforts. We are conducting a program to detect the progenitors of core-collapse SNe using a 
combination of pre-explosion {\it HST} images and post-explosion ground-based adaptive
optics (AO) imaging (e.g., Gal-Yam et al. 2005; 2006). This program is aimed at providing
additional links in the progenitor-SN map shown in Fig. 1. In parallel, we have been
working on a large study of the typical properties and fractions of core-collapse SNe of
various sub-types, in order to provide a well-defined subtype division 
(the currently used classification method
of SNe is a mix of classification by spectra and by light-curve shapes)
and a data base that can be used to interpret the results of 
large surveys at high redshift. Here we report the first results from
this effort, the Caltech Core-Collapse Project (CCCP). 

\begin{figure}
  \includegraphics[height=.5\textheight]{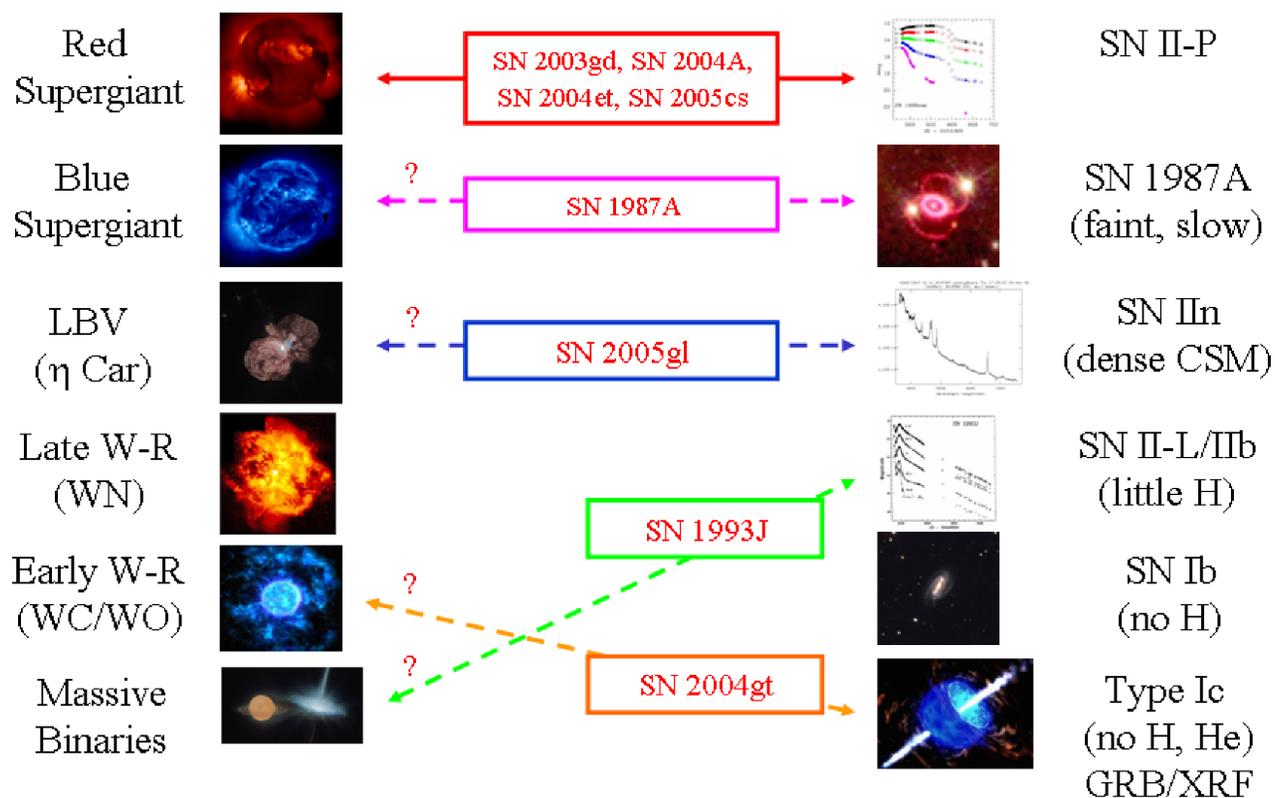}
  \caption{The progenitor-SN map based on the total census of progenitor observations (Gal-Yam et al. 2006 and references therein). The association of SNe II-P with red supergiants seems robust, all other associations
are based on single events. Additional progenitor detections (of order 1-2 per year possible) are crucial for further progress. For example, conventional theory holds that only the most massive single stars give rise to SNe Ib/c, but the theory is virtually unconstrained by observations - future progenitor identifications (or strict upper limits on the luminosity of these stars) 
should put our understanding of SNe to the test. The two most recent additions (SN 2004gt and SN 2005gl) are based on results from our AO program.}
\end{figure}

\section{The Caltech Core-Collapse Project (CCCP)}

The CCCP (\url{http://www.astro.caltech.edu/~avishay/cccp.html}) is
a large observational study of core-collapse SNe. Uniform, high-quality NIR and optical photometry
and multi-epoch optical spectroscopy (Fig. 2) have been obtained using the 200'' Hale and robotic 60''
telescopes at Palomar, for a sample of 50 nearby core-collapse SNe (Table 1). The combination of
both well-sampled optical light curves and multi-epoch spectroscopy will allow to disentangle
spectroscopically and photometrically based subtype definitions. Multi-epoch spectroscopy
is crucial to identify transition events that evolve among subtypes with time (e.g., SNe
IIb, initially similar to H-rich SNe II, but later evolving into H-poor and He-rich SNe Ib --
several such events have been identified among the CCCP sample). The CCCP SN sample
includes every core-collapse SN discovered between July 2004 and September 2005 that was visible from
Palomar, found shortly ($< 30$ days) after explosion (based on available pre-explosion
photometry), and was closer than $\sim120$ Mpc. This complete sample allows, for the first time, a
study of core-collapse SNe as a population, rather than as individual events.

\section{Preliminary Results}

\subsection{The CCCP SN sample}

In Table 1 we list the full CCCP sample of SNe. Of the 50 core-collapse events,
37 are distributed among the various type II sub-classes, and 13 events are of 
types Ib and Ic. The sample includes two events (QuestSN1 and SNF0630) discovered
using the Palomar 48'' telescope by the Quest/SN factory consortium, and 48 events
announced by various discovering groups in IAU circulars. Sample CCCP data (collected for
the three first CCCP events) are shown in Fig. 2. Unfortunately, we were not able to secure
such high-quality data for the entire CCCP sample, mainly due to the impact of the 
severe 2004-2005 California winter. However, data of similar quality exist for the majority
of events studied. As can be seen from Fig. 2, our selection criteria for young events
generally permitted us to obtain light curves that covered the epoch of peak brightness, 
and the large collecting area of the Palomar 200'' enabled us to secure high S/N spectra.  

\begin{figure}
  \includegraphics[height=.5\textheight]{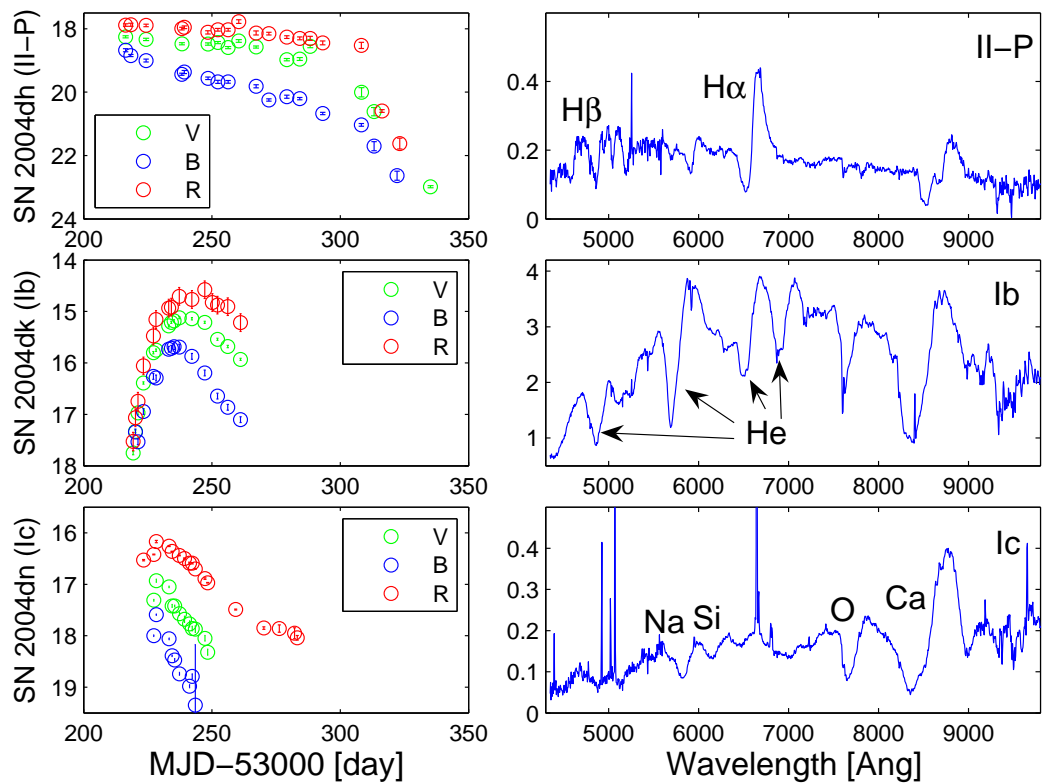}
  \caption{Example data for the first three CCCP SNe. 
Note well-sampled multi-color
light curves and high S/N spectra, combined to accurately reveal the SN sub-type. Main species are marked.
From Gal-Yam et al. 2007 (in preparation).}
\end{figure}

As part of the activity associated with the CCCP effort we have also collected data for 
a number of events that did not satisfy our strict selection criteria. These include
type II SNe 2004A (data published in Hendry et al. 2006), 2004T, 2004V and 2004Z 
observed as a pilot study of SNe II-P, observations of the bright nearby SN 2004dj
(published in Leonard et al. 2006), observations of the peculiar type IIb 
SN 2004cs (published in Rajala et al. 2005), and observations of type IIn SN 2005gl
for which we conducted a progenitor study (Gal-Yam et al. 2006). 

\begin{table}
\begin{tabular}{llllllllll}
\hline
\tablehead{1}{l}{b}{SN} &
\tablehead{1}{l}{b}{Type} &
\tablehead{1}{l}{b}{SN} &
\tablehead{1}{l}{b}{Type} &
\tablehead{1}{l}{b}{SN} &
\tablehead{1}{l}{b}{Type} &
\tablehead{1}{l}{b}{SN} &
\tablehead{1}{l}{b}{Type} &
\tablehead{1}{l}{b}{SN} &
\tablehead{1}{l}{b}{Type}   \\
\hline
2004dh & II & 2004dk & Ib & 2004dn & Ic & 2004dr & II & 2004du & II\\
2004eb & II & 2004em & II & 2004ek & II & 2004er & II & 2004et & II\\
QuestSN1 & II & 2004ex & II & 2004fc & II & 2004fe & Ic & 2004ff & Ic\\
2004fx & II & 2004ge & Ic & 2004gq & Ib/c & 2004gt & Ic & 2004gv & Ib/c\\
2004gy & II & 2005E & Ib & 2005H & II & 2005O & Ib & 2005U & II \\
2005Y & II & 2005Z & II & 2005aa & II & 2005ab & II & 2005ad & II \\
2005an & II & 2005ap & II & 2005ar & Ib & 2005au & II & 2005ay & II \\
2005az & Ib & 2005ba & II & 2005bf & Ib/c & 2005bp & II & 2005bw & II \\
2005bx & II & 2005by & II & 2005ci & II & 2005cl & II & 2005cp & II \\
2005cs & II & SNF0630 & II & 2005db & IIn & 2005dp & II & 2005ds & II \\
\hline
{\bf II} & 37 & {\bf Ib/c} & 13 & {\bf Total} & 50 & & & & \\
\hline
\end{tabular}
\caption{The CCCP Supernova sample}
\label{tab:a}
\end{table}

\subsection{SN II Demography}

Analysis of the data collected by the CCCP program is in progress, and we are currently
collecting the last observations required for field calibration and to serve as 
image-subtraction templates (only possible once the SNe have faded). A complete
analysis of the data is therefore not yet possible. Here, we present a preliminary study
of the first $\sim1/3$ of our sample, for which light curves and reduced spectra are
available, though not always fully calibrated. In particular, we have used the first
13 type II SNe for which good data exist to inspect the relative fractions of SNe II
of the various sub-types. Table 2 presents our refined typing for these events. In the table
we report both photometric type from our light curves (P = Plateau; L = rapidly declining; see
below for an explanation about the events noted as peculiar) and
the spectroscopic type (IIb = He lines appear with time; II-P = detailed analysis shows
similarity to prototypical SNe II-P such as SN 1999em, references given; II - preliminary 
inspection does not show any noteworthy features). 

As can be seen, of these first 13 events eight appear to be SNe II-P. Two events (SNe 2004ek and
2004em) show peculiar light curves dominated by late-time humps rising on a timescale of tens of days (Fig. 3).
SN 2004em appears to resemble a brighter version of SN 1987A, perhaps similar to SN 1998A 
(Pastorello et al. 2005), while SN 2004ek has an even more peculiar light curve. Regardless
of their exact light curve shape, both events certainly do not decay rapidly as SNe II-L,
and we can assume that sparser observations might have led to their classification as
SNe II-P. We therefore conclude that 8(10) of our first 13 SNe II belong to the II-P 
subclass (depending on the inclusion of these two peculiar events), two events are of type
Ib, and one (SN 2004dr) is of type II-L (which we here define as a SN with a rapidly declining
light curve which does not show He signatures in its spectra). The more rare SNe IIn are not
represented in this first sub-sample of SNe from the CCCP (though several such events are
included in the full sample; Table 1). Assuming that our sample does not suffer from strong selection 
effect (being almost volume, rather than flux, limited) we tentatively deduce that a fraction $f = 60-75\%$
of SNe II, and $\sim50\%$ of core-collapse SNe, are of type II-P. 

\begin{figure}
  \includegraphics[height=.5\textheight]{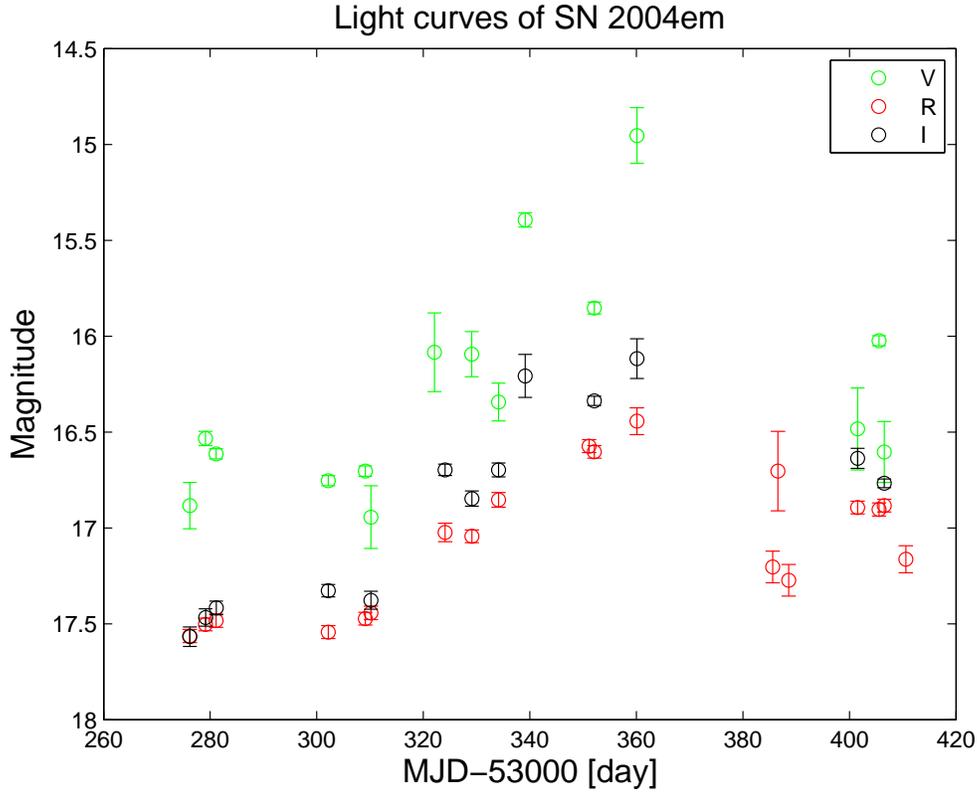}
  \caption{The peculiar light curve of SN 2004em, based on preliminary CCCP
photometry. Note the late-time hump peaking $\sim3$ months after
explosion.}
\end{figure}

Interpreting this result in the context of the progenitor-SN map (Fig. 1), which links SNe II-P
with red supergiants with an initial low main sequence mass ($8-15 {\rm M}_{\odot}$), we calculate
the fraction of such stars from the total population of core-collapse-producing stars (${\rm M}>8 {\rm M}_{\odot}$).
Assuming a Salpeter mass function, we find the the expected fraction of SNe II-P (assuming
all SNe II-P come from low-mass red supergiants, and that all stars in the above mass range 
explode as SNe II-P) is $f=57\%$. The nice agreement with our preliminary findings above
is encouraging, and may tell us that the above assumptions are correct. In particular, if most
$8-15 {\rm M}_{\odot}$ stars explode as SNe II-P, perhaps stars in this mass range, even 
if in binary systems, do not contribute significantly to the population of stripped (Ib/c)
SNe. 

\begin{table}
\begin{tabular}{llll}
\hline
\tablehead{1}{l}{b}{SN} &
\tablehead{1}{l}{b}{Photo Type} &
\tablehead{1}{l}{b}{Spec Type} &
\tablehead{1}{l}{b}{Reference} \\
\hline
2004A & P & II-P & Hendry et al. 2006 \\ 
2004cs & L & IIb & Rajala et al. 2005 \\
2004dh & P & II-P & Nugent et al. 2006 \\
2004dr & L & II & \\ 
2004du & P & II & \\ 
2004ek & P(pec) & II & \\
2004em & P(pec) & II & \\ 
2004er & P & II & \\ 
2004et & P & II-P & Li et al. 2005\\
2004ex & L & IIb & \\ 
2004fc & P & II & \\
2004fx & P & II-P & Hamuy et al. 2006\\
2005ay & P & II-P & Bufano et al. 2006\\
\hline
\end{tabular}
\caption{Type II Supernovae from the CCCP}
\label{tab:a}
\end{table}

\section{Conclusions}

We have conducted a large observational program focused on core-collapse SNe, the CCCP.
50 SNe have been observed, and data analysis is in progress. While exploiting the full potential
of this large data set will require additional time and effort, preliminary results
from the CCCP and associated projects are emerging. Here, we have shown an initial
inspection of type II SN demographics, which appear to be consistent with the idea
that SNe II-P are associated with low-mass progenitor stars.  

\begin{theacknowledgments}
A.G. acknowledges support
by NASA through Hubble Fellowship grant \#HST-HF-01158.01-A awarded by
STScI, which is operated by AURA, Inc., for NASA, under contract NAS
5-26555. A.G. further acknowledges the hospitality of the community of Cefalu and the efforts of the
organizers of the 2006 Cefalu international astronomy conference, during 
which this work has come to fruition. 
D.C.L. acknowledges support from a National Science Foundation (NSF) Astronomy and Astrophysics
Postdoctoral Fellowship (award AST-0401479), under which part of this work
was completed.
D.J.S. acknowledges support provided by NASA through Chandra Postdoctoral
Fellowship grant number PF5-60041.
\end{theacknowledgments}

\bibliographystyle{aipprocl} 

\end{document}